\documentclass{template_ECS}

\usepackage{subcaption}
\usepackage{comment}
\usepackage{dblfloatfix}
\usepackage{xr}
\usepackage[normalem]{ulem}

\makeatletter
\newcommand*{\addFileDependency}[1]{
\typeout{(#1)}
\@addtofilelist{#1}
\IfFileExists{#1}{}{\typeout{No file #1.}}
}\makeatother

\newcommand*{\myexternaldocument}[1]{%
\externaldocument{#1}%
\addFileDependency{#1.tex}%
\addFileDependency{#1.aux}%
}

\myexternaldocument{main_supp_ECS}

\addtokomafont{sectioning}{\boldmath}

\title{\raggedright Combining Molecular Dynamics and Experimental Methods for the Parametrization of Binary Carbonate-Based Electrolytes}

\author{
\begin{minipage}{\textwidth}	
	Lukas Lehnert,\textsuperscript{[1, 2]} Martin Lorenz,\textsuperscript{[3]} Maria Fernanda Juarez,\textsuperscript{[1, 2]} Max Schammer,\textsuperscript{[1, 2, 4]} Maryam Nojabaee,\textsuperscript{[5]} Monika Schönhoff,\textsuperscript{[3]} Birger Horstmann\textsuperscript{[1, 2, 6, z]}\\
	\normalsize\textit{\textsuperscript{1}German Aerospace Center, 89081 Ulm, Germany\\
	\textsuperscript{2}Helmholtz Institute Ulm, 89081 Ulm, Germany\\
	\textsuperscript{3}University of Münster, 48149 Münster, Germany\\
	\textsuperscript{4}Athene Patent, 81829 München, Germany\\
	\textsuperscript{5}German Aerospace Center, 70569 Stuttgart,
Germany\\
	\textsuperscript{6}University of Ulm, 89081 Ulm, Germany\\
	\textsuperscript{z}E-mail: birger.horstmann@dlr.de}
\end{minipage}
}

\renewcommand{\dedication}{
	\begin{minipage}{\textwidth}
	\end{minipage}
}

\renewcommand{\abstract}{Modelling the ionic transport in battery cells requires precise parametrization of the involved electrolytes. For carbonate-based electrolytes, however, the evaluation of their parameters suffers from interphase effects between the bulk electrolyte and the Li metal electrode, commonly present in the usual electrochemical polarization experiments. In this work, we combine measurements on
conductivity and concentration cells with molecular dynamic simulations, avoiding these difficulties and thus, allowing for a more accurate determination of the parameters. We determine the conductivity, the transference number, the thermodynamic factor and the salt diffusion coefficient for three different electrolytes, i.e mixtures of ethylene carbonate (EC), ethyl methyl carbonate (EMC), methyl propionate (MP), dimethyl carbonate (DMC) and propylene carbonate (PC), containing LiPF$_6$ at various concentrations and temperatures. In order to validate the simulated transference numbers, we employ electrophoretic Nuclear Magnetic Resonance spectroscopy (eNMR).
}

\begin{document}


\vspace{-1.5cm}\maketitle\vspace{-1cm}
	\textit{\dedication}\vspace{0.4cm}
\small{\begin{shaded}
		\noindent\abstract
	\end{shaded}
}



\section*{Introduction}
\label{introduction}
The optimization of electrolytes plays a major role in developing Li$^+$ ion batteries. Adjusting the electrolyte's composition allows for matching the transport properties to the battery operating conditions and can considerably improve the battery performance. This requires precise knowledge of the corresponding electrolyte parameters, and thus calls for accurate characterization methods.\\
According to the concentrated solution theory from Latz's group,\cite{Latz2011, Latz2015, Schammer2021} binary electrolytes containing three components (e.g. cations, anions, solvent) comprise four independent parameters: the conductivity $\kappa$, a transference number $t_\alpha$ (where we choose the transference number of the Li$^+$ ions $\alpha = +$ for this paper), the thermodynamic factor $\mathit{TDF}$, and the salt diffusion coefficient $D_\pm$. While electrochemical impedance spectroscopy (EIS) of the electrolyte reveals $\kappa$, measurements on concentration cells determine convoluted information about $t_+$ and $\mathit{TDF}$.\cite{Landesfeind2019,Nyman2008, Bergstrom2021, Valoen2005} 
Polarization experiments with symmetrical Li | electrolyte | Li cells enable the deconvolution of the data from the concentration cells  and the measurement of the diffusion coefficient $D_\pm$. However and in contrast to the reproducible EIS and concentration cell techniques, the polarization experiments show inconsistent results for carbonate-based electrolytes.\cite{Landesfeind2019, Nyman2008,Bergstrom2021,Lehnert2024} As discussed in Refs. \citenum{Bergstrom2021}, \citenum{Lehnert2024} and \citenum{Talian2019}, interphasial effects between the Li metal electrodes and the electrolyte interfere with the experiments, hindering the evaluation of the polarization response.\\
Electrophoretic Nuclear Magnetic Resonance spectroscopy (eNMR) provides an experimental alternative to measuring $t_+$,\cite{Gouverneur2018,Brinkkoetter2018,Gouverneur2015,Zhang2014} away from possible parasitic interphasial effects. In eNMR experiments, the electrolyte is exposed to an electric field, which induces a directed migration of the ionic species. The experiment measures the average drift velocity of each constituent in the bulk electrolyte and allows for an undisturbed, direct determination of $t_+$.\cite{Gouverneur2018,Brinkkoetter2018,Gouverneur2015,Zhang2014}  \\ 
Numerical methods like molecular dynamic (MD) simulations are completely unaffected by experimental difficulties.\cite{Fong2020a,Fong2020,Ringsby2021,Fang2023,Shim2018,Mistry2023} MD simulations model the interactions between the atoms of a particle ensemble based on the underlying applied force-fields. Calculating the corresponding individual trajectories provides a deep insight in the temporal evolution of the examined system and its equilibrium fluctuations. According to the fluctuation-dissipation theorem,\cite{Kubo1966} the statistical behavior of these microscopic fluctuations determines the macroscopic thermodynamics of the particle ensemble and thus, reveals the desired electrolyte parameters.\\
Mistry et al. developed a set of non-linear algebraic equations, connecting the relative displacements of ions and solvent molecules to the Stefan-Maxwell diffusivities, which are directly related to the transport parameters of the corresponding electrolyte. This approach requires remarkably small trajectory information compared to similar methods to predict well-behaved trends with bounded variabilities.\cite{Mistry2023}\\
Fong et al. derived Green-Kubo relations, connecting correlations of particle flux fluctuations to the Onsager transport coefficients.\cite{Fong2020a, Fong2020} While this approach might not quite match the precision of Mistry et al., it still provides reasonably accurate transport coefficients within the given statistical error bars. Furthermore, the matrix of the Onsager transport coefficients is defined as positive semi-definite, a criterion which directly ensures positive entropy production and thus, thermodynamic consistency. Therefore, we follow the approach by Fong et al.\\
The combination of MD simulations and experimental data on concentration cells allows for a complete determination of the electrolyte parameter.  Calculating the Onsager transport coefficients directly yields $\kappa$ and $t_+$. The obtained transference number $t_+$ allows for deconvoluting the experimental concentration cell data, revealing the thermodynamic factor $\mathit{TDF}$. Specifying $\mathit{TDF}$ finally enables the evaluation of the salt diffusion coefficient $D_\pm$.\\ 
The comparison of calculated conductivity values with experimental data allows for refining the MD simulations. Simulations using unpolarizable force fields often introduce an electrolyte specific scaling factor, which lowers the effective charges of the involved ions in order to compensate for the neglected screening effects due to solvent polarization.\cite{Ringsby2021,Chaban2011,Doherty2017} Benchmarking the simulated conductivity results against experimental data enables identifying the scaling factor, and thus, yields more accurate simulation results.\cite{Bedrov2019}\\

In this work, we determine the four electrolyte parameters of three systems, containing LiPF$_6$ in mixtures of ethylene carbonate (EC), ethyl methyl carbonate (EMC), methyl propionate (MP), dimethyl carbonate (DMC) and propylene carbonate (PC), at various concentrations (0.1\,M $\leq c\leq$ 3.0\,M) and temperatures ($-20$\,\textdegree C $\leq T\leq$ 20\,\textdegree C). The examined solutions are EC:EMC (3:7, weight), EC:DMC:PC (27:63:10, volume), and EC:EMC:MP (2:6:2, volume).\\
LiPF$_6$ in EC:EMC has already been examined many times, using electrochemical and numerical methods.\cite{Nyman2008,Landesfeind2019,Bergstrom2021,Ringsby2021,Logan2018} Thus, choosing this electrolyte for parametrization in our work allows for implementing experimental literature data in our characterization process, comparing the data to our results and benchmarking our numerical methods.
LiPF$_6$ in EC:DMC:PC has also been experimentally characterized,\cite{Valoen2005} which likewise benefits our study. Additionally, although with unknown solvent mass or volume ratios, this electrolyte is employed in the commercial cell LG 18650 HG2.\cite{Nguyen2019} The cell is currently under investigation in our group and thus, evaluating the parameters of the electrolyte in absence of possible, detrimental interphase effects between Li metal and the bulk electrolyte could therefore support future battery cycling simulations.
The electrolyte LiPF$_6$ in EC:EMC:MP using $c=1.0$\,M has been developed by the Jet Propulsion Laboratory for the Mars InSight mission\cite{Smart2018,Smart2008,Smart2010} and has space proven itself on the surface of Mars for several years. Therefore, this electrolyte poses an interesting candidate for its low temperature behavior.\\
For the determination of the electrolyte parameters, we combine EIS and measurements on concentration cells with MD simulations, using unpolarizable Optimized Potentials for Liquid Simulations (OPLS)\cite{Banks2005} with an optimized scaling factor for the charges of the ions. In order to validate the modeled transference numbers, we conduct additional eNMR measurements of LiPF$_6$ in EC:EMC and in EC:EMC:MP at 0.5\,M $\leq c\leq$ 1.5\,M and $T=20\,$\textdegree C. Additionally, we investigate the simulated electrolyte viscosity using the Stokes-Einstein relation, the diffusion mechanism, and the formation of ion associations, and we estimate qualitatively their influence on the transport parameters.\\

The subsequent section gives a brief overview of the theory governing the equations of the MD simulations, the EIS, the concentration cells, and the eNMR measurements. This section is followed by the evaluation of the electrolyte parameters and their fitting functions. The final section estimates the dependence of the transport parameters on the additionally calculated electrolyte characteristics mentioned above.

\section*{Theory}
\label{theory}
\subsection*{Molecular Dynamics (MD)}
\label{theory_molecular_dynamics}
We calculate the transport parameters for LiPF$_6$ in EC:EMC, in EC:DMC:PC and in EC:EMC:MP using MD simulations. For this, we summarize the essential equations in this section, following the theoretical framework derived by Fong et al. A detailed derivation is given in Refs.\citenum{Fong2020a}, \citenum{Fong2020}, and \citenum{Fong2022}.\\ 
Fong et al. base their theory on the Onsager transport equations
\begin{equation}
    \boldsymbol J_i = \sum_j L^{ij}\nabla\mu_j,
\end{equation}
where $\boldsymbol J_i$ denotes the particle flux of species $i$ relative to the center-of-mass velocity and $\mu_j$ is the electrochemical potential of species $j$. The flux and the driving forces are linearly coupled via the Onsager transport coefficients $L^{ij}$. These coefficients are later translated into the transport parameters of the electrolyte (see Eqs. \ref{theory_molecular_dynamics_kappa}\textendash \ref{theory_molecular_dynamics_D}).\\
In order to calculate $L^{ij}$, Fong et al. derive Green-Kubo relations connecting the Onsager coefficients to a correlation function between the fluxes $\boldsymbol J_i$ and $\boldsymbol J_j$,
\begin{equation}
\label{Lij_formula_1}
    L^{ij} = \frac{V}{3k_\mathrm{B}T}\int_{0}^{\infty} \mathrm{d}t\langle \boldsymbol J_i\left(t\right) \cdot\boldsymbol J_j\left(0\right) \rangle
\end{equation}
with the ensemble volume $V$, the Boltzmann constant $k_\mathrm{B}$, and the temperature $T$. Noticeable this equation suggests interpreting $L^{ij}$ as a degree of motion correlation between the species $i$ and $j$.\cite{Fong2020} The equation can be also reformulated into an expression depending on the individual particle positions $\boldsymbol r_i^\alpha$ and $\boldsymbol r_j^\beta$ with respect to the position of the center-of-mass,\cite{Fong2020}
\begin{equation}\begin{aligned}
\label{Lij_correlation_equation}
    L^{ij} &= \frac{1}{6k_\mathrm{B}TV} \lim_{t \to \infty} \frac{\mathrm{d}}{\mathrm{d}t} \biggl< \sum_{\alpha} \left[ \boldsymbol r_i^\alpha\left(t\right) - \boldsymbol r_i^\alpha\left(0\right) \right]\\
    &\quad \cdot \sum_{\beta} \left[ \boldsymbol r_j^\beta\left(t\right) - \boldsymbol r_j^\beta\left(0\right) \right] \biggl>.
    \end{aligned}
\end{equation}
Here, we use equation \ref{Lij_correlation_equation} to determine the coefficients $L^{ij}$ for binary electrolytes with three components: Li$^+$, and PF$_6^-$ ions, and solvent molecules. The number of the corresponding independent Onsager transport coefficients reduces to $n(n-1)/2=3$ (where $n$ is the number of components), due to the Onsager reciprocal relations $L^{ij} = L^{ji}$ .\cite{Onsager1931, Onsager1931a} Therefore, we evaluate the particle trajectories of the ions to calculate $L^{++}$, $L^{--}$ and $L^{+-}$. Note that this merging of multiple solvent species into one single "pseudo" solvent simplifies the underlying transport mechanisms. As previous research shows, the components of a solvent mixture exhibit their individual mobilities, contributing to the transport phenomena in the corresponding electrolyte.\cite{Steinrueck2020, Wang2022,Mistry2022, Jung2023,Mistry2025} Thus, a complete and detailed description of the transport properties would require the determination of an increased number of Onsager transport coefficients, accounting for these mobilities. However, we believe that an accurate electrochemical battery model can be still achieved even using a simplified model with lower computational complexity. The advantages of using these valid simplifications relies in the possibility of describing with higher precision the internal electrochemical mechanism of batteries. In these cases, other advanced techniques, like machine learning or quantum computing algorithms, can be applied to improve the accuracy of the models. The use of these new technologies allows also to increase the complexity of the initial models, as stacking learning ensembles can be used in these algorithms.\\
The Onsager transport coefficients are translated into the transport parameters of the electrolyte. For our binary systems (with the stoichiometric factors $\nu_+=\nu_-=1$) the conductivity $\kappa$, the transference number $t_+$ and the salt diffusion coefficient $D_\pm$ read\cite{Fong2020a,Fong2020, Fong2022}
\begin{equation}
\label{theory_molecular_dynamics_kappa}
    \kappa = F^2\left(z_+^2L^{++} + 2z_+z_-L^{+-} + z_-^2L^{--}\right),
\end{equation}
\begin{equation}
\label{theory_molecular_dynamics_t+}
    t_+ = \frac{z_+^2L^{++} + z_+z_-L^{+-}}{z_+^2L^{++} + 2z_+z_-L^{+-} + z_-^2L^{--}},
\end{equation}
\begin{equation}
\label{theory_molecular_dynamics_D}
    D_\pm = \frac{-z_+z_-\left(L^{++}L^{--}-{L^{+-}}^2\right)}{z_+^2L^{++} + 2z_+z_-L^{+-} + z_-^2L^{--}}\frac{RT}{c}\mathit{TDF}
\end{equation}
with the charge numbers $z_+=1$ and $z_-=-1$ of the ions, the Faraday constant $F$, and the ideal gas constant $R$. The thermodynamic factor $\mathit{TDF} = 1+\frac{\mathrm{d}\ln{f_\pm}}{\mathrm{d}\ln{c}}$ comprises the salt activity coefficient $f_\pm$ and is generally unknown. In order to reveal $\mathit{TDF}$, the authors combine the MD simulations with concentration cell measurements, yielding a convoluted expression of $t_+$ and $\mathit{TDF}$ as a function of $c$ and $T$ (see Eqs. \ref{eq_U_c}\textendash \ref{eq_TDF_decon} in the subsequent section). The combination allows for deconvoluting the thermodynamic factor $\mathit{TDF}$ and hence enables the calculation of $D_\pm$, completing the determination of the four electrolyte parameters. Alternatively, although not done in this work, the $\mathit{TDF}$ can be directly calculated from MD simulations using numerical methods such as thermodynamic integration, Kirkwood-Buff integrals or the S0 method.\cite{ Sanz2007, Paluch2010, Kirkwood1951, Dawass2019, Cheng2022} Note that Fong et al. define $\mathit{TDF}$ using the chemical potential of the Li$^+$ ions.\cite{Newman2004} Instead, we use the chemical potential of the neutral salt summing up the chemical potential of anions and cations, weighted with the correct stoichiometric factors.\cite{Latz2011, Latz2015} Therefore, equation \ref{theory_molecular_dynamics_D} deviates by a factor $\frac{1}{2}$ from the relation in Ref.\citenum{Fong2020a}. Technical details on the employed simulations and the computational methods can be found in the supplementary information (see SI Section S1.1\textendash S1.4).\cite{Ringsby2021, Thompson2022, Banks2005, Schrodinger2021, Jensen2006, CanongiaLopes2004, Chaban2011, Martinez2009, Nose1984, Nose1984a, Hoover1985, Hoover1986, Swope1982, Hockney1989, Fong2020, Fong_git, Michaud‐Agrawal2011, Gowers2016, Sigma, Kowsari2011, Savitzky1964, Valoen2005, Hall2015, Naejus1997, Landesfeind2019}\\
Considering our transport theory, the fluxes of the individual species are defined with respect to some reference velocity $\boldsymbol v^\Psi$.\cite{Schammer2023} The choice of the reference frame $\Psi$ directly affects the effective electrochemical potential of the electrolyte $\mu^\Psi$ and the Onsager transport coefficients $L^{ij,\Psi}$. Therefore, MD simulations and experiments conducted in different reference frames yield deviating electrolyte parameters. A direct comparison of the parameters thus requires a transformation into the same reference frame.\\
In MD, the calculation of the transport parameter (see Eqs. \ref{theory_molecular_dynamics_kappa}\textendash \ref{theory_molecular_dynamics_D}) often uses as reference frame the velocity of the center-of-mass (COM).\cite{Ringsby2021} In contrast, some other groups have shown, that eNMR measurements capture the transference number $t_+$ in a reference frame with local volume conservation (VOL). \cite{Lorenz2022, Kilchert2023} In order to compare the MD simulation to the eNMR experiments, we transform the calculated parameters from the COM into the VOL reference frame (see SI Eqs. S6\textendash S8).\cite{Schammer2023, Lundgren2014, Berhaut2015} For our local electro-neutral electrolytes, the transformations read
\begin{equation}
\kappa^\mathrm{VOL} = \kappa^\mathrm{COM},
\end{equation}
\begin{equation}
t_+^\mathrm{VOL} = \frac{c_0\nu_0}{\omega_0}\left(t_+^\mathrm{COM} - \omega_-\right) + c_-\nu_-,
\end{equation}
\begin{equation}
\frac{D_\pm^\mathrm{VOL}}{\mathit{TDF}^\mathrm{VOL}} = \frac{c_0^2\nu_0^2}{\omega_0^2}\frac{D_\pm^\mathrm{COM}}{\mathit{TDF}^\mathrm{COM}}
\end{equation}
with the partial molar volume $\nu_i$ and the mass fraction $\omega_i$ of species $i$. Based on the findings of Lorenz \cite{Lorenz2022} and Kilchert \cite{Kilchert2023}, we expect that the concentration cells capture information about the transference number $t_+$ and the thermodynamic factor $\mathit{TDF}$ in the volume-based frame (see Eqs. \ref{eq_U_c}\textendash \ref{eq_TDF_decon}).\\
While we focus on the volume-based frame in this work, the literature often presents the electrolyte parameters in different reference frames. Therefore, we additionally present our findings in the COM and in the solvent-velocity reference frame (SOL) in the supplementary information (see SI Figures S3, S4 and SI Tables S5, S6).

\subsection*{Experiments}
\hspace*{1em}\textit{Conductivity Cell}\\
Experimental conductivity data enable benchmarking and improving the MD simulations by adjusting the scaling factor for the effective charges of the involved ions. Performing EIS reveals the bulk resistance $R_\mathrm{el}$ of the electrolyte, which translates to the corresponding conductivity $\kappa$ with\cite{Landesfeind2019,Nyman2008, Bergstrom2021, Valoen2005}
\begin{equation}
\kappa = \frac{k}{R_\mathrm{el}}.
\end{equation}
The cell constant $k$ captures the geometrical characteristics of the conductivity cell and can be measured by using a reference electrolyte with known conductivity. Experimental details can be found in the supplementary information (see SI Section S2.1).\\

\textit{Concentration Cell}\\
Concentration cells comprise two identical electrodes (in our case Li metal electrodes), immersed in separated electrolyte solutions with slightly different concentrations $c_\mathrm{L}$ and $c_\mathrm{R}$. A salt bridge ionically connects the two solutions, enabling a finite but negligibly small current to flow. This allows for measuring the concentration potential $U_\mathrm{c}$ between the electrodes. Considering the theory of concentrated electrolytes introduced by Latz et al.,\cite{Latz2011, Latz2015}, $U_\mathrm{c}$ includes convoluted information about the transference number $t_+$ and the thermodynamic factor $\mathit{TDF}$ in form of $a(c, T) = \left( 1-t_+\right)\mathit{TDF}$,
\begin{equation}
\label{eq_U_c}
U_\mathrm{c} = \frac{RT}{F}\int_{c_\mathrm{L}}^{c_\mathrm{R}} a(c,T)\,d\ln{c}.
\end{equation} 
Following the approach of Val\o en et al.,\cite{Valoen2005} $a(c, T)$ can be expressed by a Taylor expansion, 
\begin{equation}\label{eq_a(c,T)}\begin{aligned}
a(c,T) &= a_0(T) + a_1(T)c^{1/2} + a_2(T)c^1 + \cdots \\
&\quad = \sum_{i=0}^n a_i(T)c^{i/2},
\end{aligned}
\end{equation}
where the coefficients $a_i(T)=a_{i0}\left[1+a_{i1}\left(T-T_0\right)\right]$ depend linearly on the temperature with $T_0=293.15\,$K.
The insertion of equation \ref{eq_a(c,T)} into equation \ref{eq_U_c} and subsequent integration relates the potential $U_\mathrm{c}$ directly to the coefficients $a_i(T)$,\cite{Valoen2005}
\begin{equation}
\label{eq_U_c_int}
\frac{F}{RT}U_\mathrm{c} = a_0(T)\left[\ln{c}\right]_{c_\mathrm{L}}^{c_\mathrm{R}} + \sum_{i=1}^n a_i(T) \frac{2}{i} \left[c^{i/2}\right]_{c_\mathrm{L}}^{c_\mathrm{R}}.
\end{equation}
Therefore, the measurement of multiple concentration cells at different temperatures $T$ and concentration pairs $c_\mathrm{L}$ and $c_\mathrm{R}$ allows for determining the coefficients $a_i(T)$ and thus, reveals $a(c, T)$. As mentioned above, the expressions for $U_\mathrm{c}$ derived in equations \ref{eq_U_c} and \ref{eq_U_c_int} using our theory differs by a factor of 2 from similar expressions, using the concentrated solution theory by Newman.\cite{Newman2004}\\
Combining the results of the MD simulations for $t_+$ (see Eq. \ref{theory_molecular_dynamics_t+}) with $a(c, T)$ deconvolutes the thermodynamic factor,
\begin{equation}
\label{eq_TDF_decon}
    \mathit{TDF} = \frac{a(c, T)}{1-t_+}.
\end{equation}
As mentioned above, we use the measurements on  concentration cells to determine $a^\mathrm{VOL}(c, T)$ in the VOL reference frame. For the sake of comparison and completeness, the deconvolution of $\mathit{TDF}^\mathrm{VOL}$ has therefore to be done with $t_+^\mathrm{VOL}$. Experimental details can be found in the supplementary information (see SI Sections S2.1 and S2.2).\\

\textit{eNMR}\\
Electrophoretic Nuclear Magnetic Resonance (eNMR) allows for the determination of ion-specific drift velocities $v$ in an applied electric field $E=\frac{U}{d}$. Here, $U$ denotes the voltage between the two electrodes and $d$ the electrode distance. In contrast to random diffusive motion, which leads to a decay of the measured NMR echo intensity in an echo experiment with pulsed magnetic field gradients, the coherent migrational flux of ions causes a phase shift of the NMR-signal. This can be read out using phase sensitive Lorentzian-type fit functions as previously described.\cite{Schmidt2020} The obtained phase shift $\phi-\phi_0$ is then connected with the electrophoretic mobility $u=\frac{v}{E}$ of the respective ionic constituent according to\cite{Holz1994}
\begin{equation}
\label{eq_eNMR}
\phi-\phi_0 = \delta\Delta\gamma gEu
\end{equation}
with the observation time $\Delta$, the gradient pulse duration $\delta$, the gyromagnetic ratio $\gamma$, and the gradient strength $g$. Consequently, $u$ can be determined from a linear fit of $\phi-\phi_0$ against $U$ (see SI Figure S27). The obtained electrophoretic mobilities $u_+$ and $u_-$ for the Li$^+$ and PF$_6^-$ions translate to the transference number $t_+$ according to
\begin{equation}
\label{eq_t+_eNMR}
t_+ = \frac{z_+u_+}{z_+u_+ + z_-u_-}.
\end{equation}
As mentioned above, the eNMR measurements capture the electrophoretic mobilities and transference numbers in the VOL reference frame. Experimental details can be found in the supplementary information (see SI Sections S2.1 and S2.3).\cite{Gouverneur2015, Pettersson2004}

\section*{Results}
\label{results}
This section presents the evaluation of the electrolyte parameters in the VOL reference frame for our three electrolytes at various concentrations and temperatures. The transformed parameters in the COM and SOL frame can be found in the supplementary information (see SI Figures S3, S4 and SI Tables S5, S6). While these parameter-sets are obtained using the method of Fong et al.\cite{Fong2020a, Fong2020}, we additionally apply the method of Misty et al.\cite{Mistry2023} exemplary for one electrolyte in the supplementary information (see SI Figure S7). Note that capturing crystal nucleation in liquids is challenging for MD simulations.\cite{Sosso2016} Therefore, the simulations do not predict any possible phase transitions and thus, the parameters at low temperatures have to be treated with care.  

\begin{figure}[b!]
\begin{center}
\includegraphics[width=\linewidth]{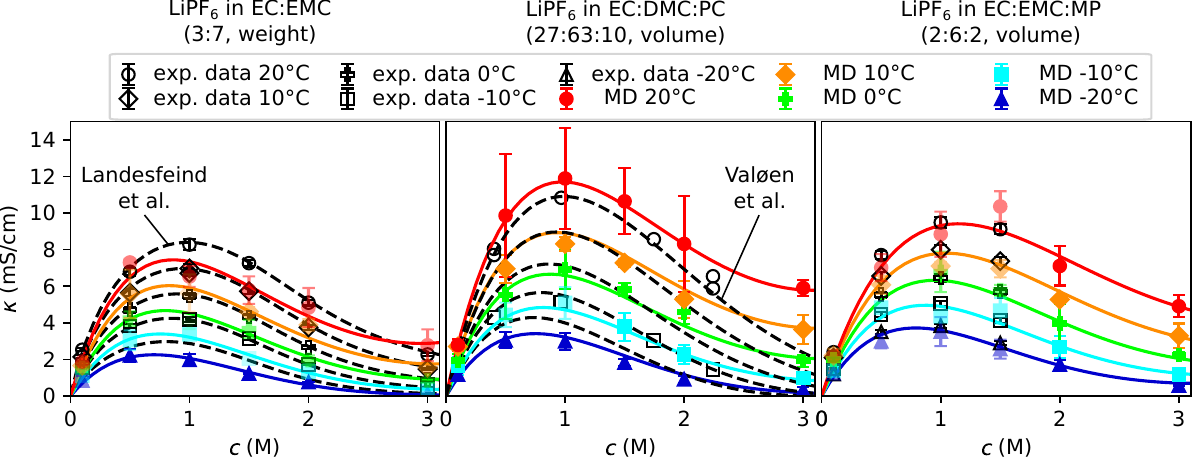}
\caption{Conductivities $\kappa$ obtained by MD simulations and electrochemical measurements, together with their corresponding fit functions (solid and dashed lines). The experimental conductivities for LiPF$_6$ in EC:EMC and in EC:DMC:PC were taken from Landesfeind et al.\cite{Landesfeind2019} and Val\o en et al.\cite{Valoen2005}}
\label{pic_cond_VOL}
\end{center}
\end{figure}
\begin{table}[b!]
	\begin{center}
	\caption{Fit coefficients $\kappa_{ij}$ (in $\left(\mathrm{mS}/\mathrm{cm}\right)^{0.5}\,\mathrm{M}^{-i-0.5}\,\mathrm{K}^{-j}$) for the conductivities $\kappa$.}
	\label{tab_cond_kappas}
		\begin{tabular}{cccc}	
\toprule		
Fit & LiPF$_6$ in EC:EMC & LiPF$_6$ in EC:DMC:PC  & LiPF$_6$ in EC:EMC:MP  \\
coefficient & (3:7, weight) & (27:63:10, volume) & (2:6:2, volume)\\
\midrule
$\kappa_{00}$ & $-2.438\cdot 10^{1}$ & $1.876\cdot 10^{0}$ & $-1.992\cdot 10^{1}$\\
$\kappa_{01}$ & $1.597\cdot 10^{-1}$ & $-3.655\cdot 10^{-2}$ & $1.471\cdot 10^{-1}$\\
$\kappa_{02}$ & $-2.067\cdot 10^{-4}$ & $1.673\cdot 10^{-4}$ & $-2.168\cdot 10^{-4}$\\
$\kappa_{10}$ & $8.677\cdot 10^{0}$ & $-2.592\cdot 10^{0}$ & $6.400\cdot 10^{0}$\\
$\kappa_{11}$ & $-6.024\cdot 10^{-2}$ & $2.035\cdot 10^{-2}$ & $-6.206\cdot 10^{-2}$\\
$\kappa_{12}$ & $7.730\cdot 10^{-5}$ & $-6.821\cdot 10^{-5}$ & $1.171\cdot 10^{-4}$\\
$\kappa_{20}$ & $-5.380\cdot 10^{-1}$ & $-4.285\cdot 10^{-1}$ & $6.877\cdot 10^{-1}$\\
$\kappa_{21}$ & $3.093\cdot 10^{-3}$ & $2.717\cdot 10^{-3}$ & $-1.628\cdot 10^{-3}$\\
\bottomrule	
	\end{tabular}
	\end{center}
\end{table}
\subsection*{Conductivity}
Figure \ref{pic_cond_VOL} shows the conductivity results $\kappa$ of the MD simulations for our three electrolytes together with the corresponding experimental data.\cite{Landesfeind2019,Valoen2005}. In general, the simulations match the experimental values well and exhibit for each electrolyte a similar trend: $\kappa$ increases with concentration up to a maximum, before it declines to lower values. Elevating the temperature enhances the conductivity.\\
The accordance to the experimental data indicates that the MD simulations capture the behaviour of the electrolytes reasonably well for concentrations below $c=2.0\,$M. However, the MD simulations overestimate the conductivity for LiPF$_6$ in EC:EMC and in EC:DMC:PC at $c=3.0$\,M. This could originate from using a constant scaling factor $\zeta$, omitting any concentration dependence of the solvent screening effects due to solvent polarization. At elevated concentrations, the ratio of ions to solvent molecules increases drastically, minimizing the amount of solvent molecules in the first solvation shell (see SI Figures S10, S11) and the corresponding solvent screening effects. Therefore, employing a constant scaling factor might underestimate the effective ionic charges at $c=3.0\,$M, resulting in an underestimated amount of ion associations and thus, yielding elevated conductivities in the MD simulations. Note that the deviation in conductivity might hint to similar deviations, occurring in other electrolyte parameters at elevated concentrations.\\
Using the functional form proposed by Val\o en et al.\cite{Valoen2005}, we fit the simulated conductivities $\kappa$ with equation \ref{eq_fit_kappa}. Table \ref{tab_cond_kappas} shows the corresponding coefficients $\kappa_{ij}$ for our three electrolytes.
\begin{equation}
\label{eq_fit_kappa}
\sqrt{\frac{\kappa}{c}} = \sum_{i=0}^n\sum_{j=0}^k\kappa_{ij}c^iT^j
\end{equation}

\subsection*{Transference Number}
\label{res_t_+}
\begin{figure}[b!]
\begin{center}
\includegraphics[width=\linewidth]{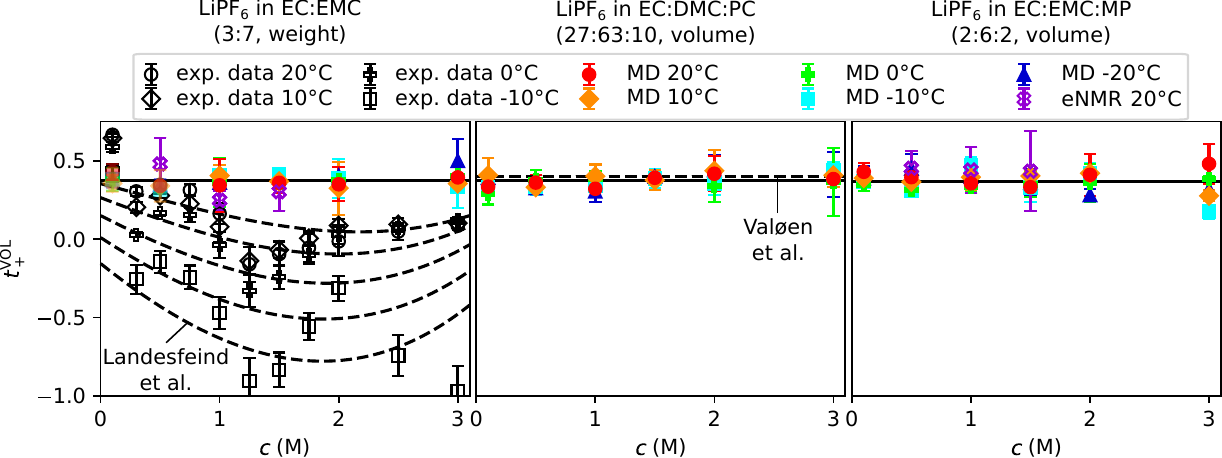}
\caption{Transference numbers $t^\mathrm{VOL}_+$ in the VOL reference frame obtained by MD simulations, electrochemical experiments and eNMR measurements, together with their corresponding fit functions (solid and dashed lines). The transference numbers of LiPF$_6$ in EC:EMC and in EC:DMC:PC were taken from Landesfeind et al.\cite{Landesfeind2019} and Val\o en et al.\cite{Valoen2005}}
\label{pic_t+_VOL}
\end{center}
\end{figure}

The MD simulations yield similar low transference numbers $t^\mathrm{VOL}_+$ for our three electrolytes (see Figure \ref{pic_t+_VOL}), indicating that primarily the PF$_6^-$ ions contribute to the ionic conductivity. This matches fairly well with the experimental results of Val\o en et al.\cite{Valoen2005} for LiPF$_6$ in EC:DMC:PC and the eNMR measurements at 20\,\textdegree C for LiPF$_6$ in EC:EMC and EC:EMC:MP. However, our findings deviate from the galvanostatic polarization data from Landesfeind et al.\cite{Landesfeind2019} for LiPF$_6$ in EC:EMC. In their work, they find that $t^\mathrm{VOL}_+$ correlates with both, concentration and temperature, yielding even negative transference numbers at low temperatures. In contrast, our MD simulations do not resolve any visible dependence of $t^\mathrm{VOL}_+$ on the concentration nor the temperature, which is consistent with the MD simulation results from Ringsby et al.\cite{Ringsby2021} at $c=1.0\,$M. The discrepancy probably originates from the difficulty of determining the transport parameters of carbonate-based electrolytes using polarization experiments with Li metal electrodes. In these experiments, porous mossy Li and solid-electrolyte interphase (SEI) layers cover the initially pristine Li metal electrode, influencing the induced concentration gradient and thus, the potential response.\cite{Talian2019, Bergstrom2021, Lehnert2024} The MD simulations as well as the eNMR measurements do not suffer from these surface effects and can therefore provide more precise values for the transference numbers $t^\mathrm{VOL}_+$. Hence, the accordance of the calculated MD data with the eNMR results validates the integrity of our simulations.\\
Since $t^\mathrm{VOL}_+$ lacks any visible trend in our MD simulations due to the scattering of the data, we fit the data with a constant. Table \ref{tab_t+} lists the corresponding values.
\begin{table}[b!]
	\begin{center}
	\caption{Constant fit coefficients for the transference numbers $t^\mathrm{VOL}_+$ from our MD simulations, compared to our eNMR measurements and the findings of Val\o en et al.\cite{Valoen2005}}
	\label{tab_t+}
		\begin{tabular}{cccc}	
\toprule		
\multirow{2}{*}{Parameter} & LiPF$_6$ in EC:EMC & LiPF$_6$ in EC:DMC:PC  & LiPF$_6$ in EC:EMC:MP  \\
& (3:7, weight) & (27:63:10, volume) & (2:6:2, volume)\\
\midrule
$t^\mathrm{VOL}_+$ (MD) & $0.373$ & $0.372$ & $0.366$\\ 
$t^\mathrm{VOL}_+$ (eNMR) & $0.343$ & - & $0.441$\\
$t^\mathrm{VOL}_+$ (Val\o en et al.\cite{Valoen2005}) & - & $0.399$ & -\\
\bottomrule	
	\end{tabular}
	\end{center}
\end{table}

\subsection*{Thermodynamic Factor}
\begin{figure}[b!]
\begin{center}
\includegraphics[width=\linewidth]{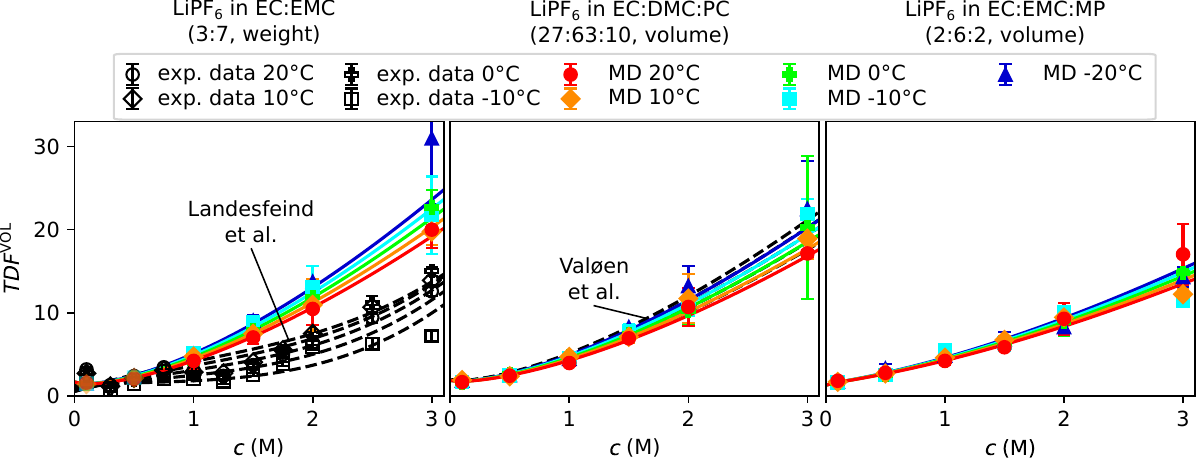}
\caption{Thermodynamic factors $\mathit{TDF}^\mathrm{VOL}$ in the VOL reference frame determined by combining concentration cell measurements with MD simulations or polarization experiments\cite{Landesfeind2019, Valoen2005}. The solid and dashed lines show their corresponding fit functions. Note that we multiply the literature data by a factor of 2, as mentioned in the main text.}
\label{pic_tdf_VOL}
\end{center}
\end{figure}

The experiments on concentration cells allow for the determination of $a^\mathrm{VOL}(c,T)=(1-t^\mathrm{VOL}_+)\mathit{TDF}^\mathrm{VOL}$, containing convoluted information about the transference number and the thermodynamic factor (see Eqs. \ref{eq_U_c}\textendash \ref{eq_U_c_int}). Hence, the combination of the MD simulations determining $t^\mathrm{VOL}_+$ and the measurements on concentration cells reveals the thermodynamic factor $\mathit{TDF}^\mathrm{VOL}$ (see Eq. \ref{eq_TDF_decon}).\\
Val\o en et al.\cite{Valoen2005} conduct experiments on concentration cells for LiPF$_6$ in EC:DMC:PC and fit the determined potentials with equation \ref{eq_U_c_int} to obtain $a^\mathrm{VOL}(c,T)$. In the fitting process, the authors set $a_2=0$ and introduce the temperature dependence solely in $a_3(T)$ while keeping $a_0$ and $a_1$ constant (i.e. $a_{01} = a_{11} = 0$). This results in a total of four fitting coefficients $a_{00}$, $a_{10}$, $a_{30}$ and $a_{31}$.\\
In order to find $a^\mathrm{VOL}(c,T)$ for LiPF$_6$ in EC:EMC and LiPF$_6$ in EC:EMC:MP, we apply the same procedure to the data provided by Landesfeind et al.\cite{Landesfeind2019} and our own concentration cell data (see SI Figure S23). Table \ref{tab_a_ij} lists the corresponding coefficients $a_i(T)$.\\
Inserting the obtained fit function $a^\mathrm{VOL}(c,T)$ together with our MD simulation results for $t^\mathrm{VOL}_+$ in equation \ref{eq_TDF_decon} determines the thermodynamic factor $\mathit{TDF}^\mathrm{VOL}$ (see Figure \ref{pic_tdf_VOL}). $\mathit{TDF}^\mathrm{VOL}$ exhibits for all three electrolytes a similar shape with increasing values with increasing concentration and decreasing temperature. While our simulated data match well with the experimental findings of Val\o en et al. for LiPF$_6$ in EC:DMC:PC, they deviate from the results from Landesfeind et al. for LiPF$_6$ in EC:EMC. The reason for this discrepancy probably arises from the challenging determination of $t^\mathrm{VOL}_+$ using Li metal electrodes together with carbonate-based electrolytes in polarization experiments, as mentioned above.\cite{Lehnert2024}\\
Note that $\mathit{TDF}^\mathrm{VOL}$ in our theory\cite{Schammer2021} differs from $\mathit{TDF}^\mathrm{VOL}$ derived by Newman's concentrated solution theory\cite{Newman2004} by a factor of $2$, as mentioned above. In order to compare with our set of fit parameters, we multiplied the literature data in Figure \ref{pic_tdf_VOL} by this factor.\\
Also note that concentration cells at very low temperatures show irregular ice formation for identical cells, revealing the electrolyte as supercooled fluid (see SI Figures S24\textendash S26).

\subsection*{Diffusion Coefficient}
\begin{table}[b!]
	\begin{center}
	\caption{Fit coefficients $T_0$ (in K) and $a_{ij}$ (in M$^{-i/2}$\,K$^{-j}$) for $a^\mathrm{VOL}\left(c,T\right)$.}
	\label{tab_a_ij}
		\begin{tabular}{cccc}	
\toprule		
\multirow{2}{*}{Parameter} & LiPF$_6$ in EC:EMC & LiPF$_6$ in EC:DMC:PC  & LiPF$_6$ in EC:EMC:MP  \\
& (3:7, weight) & (27:63:10, volume) & (2:6:2, volume)\\
\midrule
$T_0$ & 293.15 & 293.15* & 293.15 \\
$a_{00}$ & $1.085\cdot 10^{0}$ & $1.202\cdot 10^{0}$* & $8.134\cdot 10^{-1}$ \\
$a_{10}$ & $-7.761\cdot 10^{-1}$ & $-4.800\cdot 10^{-1}$* & $5.980\cdot 10^{-1}$ \\
$a_{30}$ & $2.365\cdot 10^{0}$ & $1.964\cdot 10^{0}$* & $1.291\cdot 10^{0}$ \\
$a_{31}$ & $-5.627\cdot 10^{-3}$ & $-5.200\cdot 10^{-3}$* & $-4.308\cdot 10^{-3}$ \\
\bottomrule
        \end{tabular}
    \end{center}
\footnotesize{\textsf{* Adapted from Val\o en et al.\cite{Valoen2005}}}
\end{table}
\begin{figure}[b!]
\begin{center}
\includegraphics[width=\linewidth]{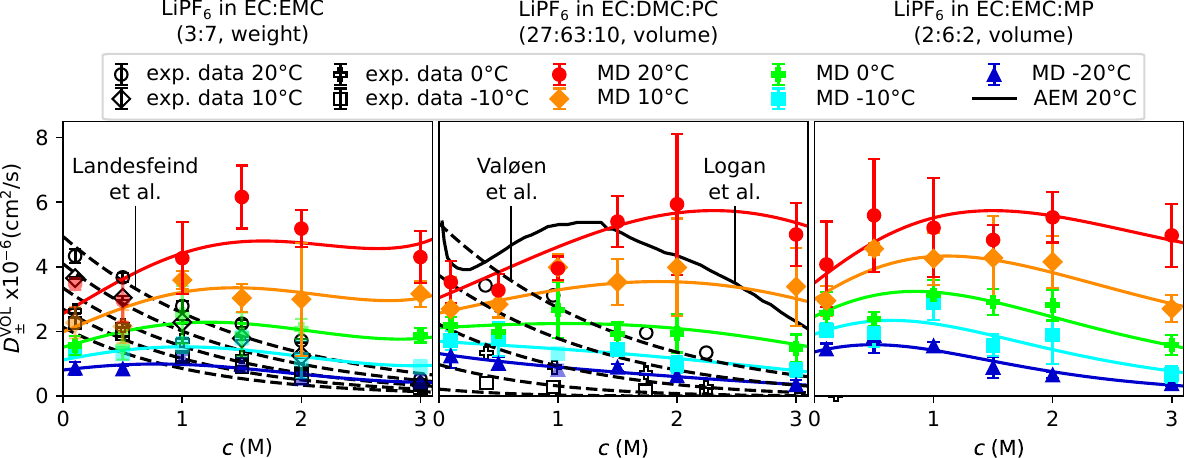}
\caption{Diffusion coefficients $D^\mathrm{VOL}_\pm$ in the VOL reference frame calculated with MD simulations, the Advanced Electrolyte Model (AEM)\cite{Logan2019, Gering2006, Gering2017} or electrochemically measured\cite{Landesfeind2019, Valoen2005} together with their corresponding fit functions. Note that the AEM calculations determine the diffusion coefficient of a similar electrolyte LiPF$_6$ in EC:DMC (3:7, weight) over molal concentrations, slightly deviating from molar concentrations (see SI Figure S14).}
\label{pic_diff_VOL}
\end{center}
\end{figure}
\begin{table}[b!]
    \begin{center}
    \caption{Fit coefficients $T_{\mathrm{g}i}$ (in K\,M$^{-i}$) and $D_{ij}$ (in K$^j$\,M$^{-i}$) for the diffusion coefficients $D^\mathrm{VOL}_\pm$.}
    \label{tab_D_ij}
        \begin{tabular}{cccc}
\toprule
\multirow{2}{*}{Parameter} & LiPF$_6$ in EC:EMC & LiPF$_6$ in EC:DMC:PC  & LiPF$_6$ in EC:EMC:MP  \\
& (3:7, weight) & (27:63:10, volume) & (2:6:2, volume)\\
\midrule
$T_{\mathrm{g}0}$ & $4.988\cdot 10^{1}$ & $1.209\cdot 10^{2}$ & $1.929\cdot 10^{2}$ \\
$T_{\mathrm{g}1}$ & $3.945\cdot 10^{1}$ & $-1.363\cdot 10^{1}$ & $-1.649\cdot 10^{1}$ \\
$D_{00}$ & $-3.047\cdot 10^{0}$ & $-4.313\cdot 10^{0}$ & $-4.840\cdot 10^{0}$ \\
$D_{01}$ & $-6.191\cdot 10^{2}$ & $-2.074\cdot 10^{2}$ & $-6.168\cdot 10^{1}$ \\
$D_{10}$ & $5.685\cdot 10^{-1}$ & $1.676\cdot 10^{0}$ & $6.214\cdot 10^{-1}$ \\
$D_{11}$ & $5.253\cdot 10^{1}$ & $-2.676\cdot 10^{2}$ & $-3.685\cdot 10^{1}$ \\
$D_{20}$ & $-1.962\cdot 10^{-1}$ & $-1.000\cdot 10^{-1}$ & $9.174\cdot 10^{-2}$ \\
$D_{21}$ & $2.050\cdot 10^{1}$ & $-8.239\cdot 10^{0}$ & $-3.354\cdot 10^{1}$ \\
\bottomrule
        \end{tabular}
    \end{center}
\end{table}
The salt diffusion coefficient $D^\mathrm{VOL}_\pm$ obtained by our MD simulations shows a clear deviation from the results of the electrochemical experiments (see Figure \ref{pic_diff_VOL}).\cite{Valoen2005, Landesfeind2019} Even though the increasing trend with temperature coincides, the concentration dependence differs and rather resembles the findings of Logan et al.\cite{Logan2019}, using the Advanced Electrolyte Model\cite{Gering2006,Gering2017} to calculate salt diffusion coefficient for LiPF$_6$ in EC:DMC (3:7, weight) at 20\,\textdegree C. While our calculations exhibit a more complex shape with elevated $D^\mathrm{VOL}_\pm$ even at higher concentrations, the experiments find monotonously decreasing values with increasing concentration. As mentioned above, this deviation could originate from additional porous layers of mossy Li and SEI, influencing the potential response in the galvanostatic polarization experiments and hence, making the electrochemical determination of the diffusion coefficient $D^\mathrm{VOL}_\pm$ difficult. We further analyze the diffusion coefficient trends in the SI (see SI Section S1.6), using LiPF$_6$ in EC:EMC as an example, and compare them to the diffusion coefficient trends of the solid polymer electrolyte lithium bis(trifluoromethanesulfonyl)imide (LiTFSI) in poly(ethylene oxide) (PEO).\cite{Gao2021} The polymer electrolyte does not suffer from detrimental effects in the polarization measurements and thus, should yield more trustworthy experimental data of the diffusion coefficient than carbonate-based electrolytes.\\
In order to represent the MD data in an analytic form, we take the same approach as Val\o en et al.\cite{Valoen2005} and fit the data to an exponential function
\begin{equation}
\log_{10} D^\mathrm{VOL}_\pm = \sum_{i=0}^nD_i(c,T)c^i,
\end{equation}
where 
\begin{equation}
D_i(c,T) = \sum_{j=0}^n \frac{D_{ij}}{\left[T-\left(T_{\mathrm{g}0} + cT_{\mathrm{g}1}\right)\right]^j}.
\end{equation}
Table \ref{tab_D_ij} shows the corresponding coefficients $T_{\mathrm{g}i}$ and $D_{ij}$ for our three electrolytes.

\section*{Discussion}
\label{discussion}
Ionic transport in concentrated electrolytes is a complex phenomenon. In order to precisely predict the electrolyte characteristics, a multitude of interactions have to be considered in detail.\cite{Gering2017} In this section, we roughly analyze the trends of the solvation shell size, the viscosity, the diffusion mechanisms, and the amount of ion association for our three electrolytes and qualitatively estimate their impacts on the ionic transport. For simplicity, we approximate and analyze the transport parameters using dilute solution theory in the COM frame. Computational details and further analysis of the radial distribution function (RDF), the coordination numbers of the ions, the viscosity, the residence time, the diffusion length, the ion association and the ionic mobility can be found in the supplementary information (see SI Section S1.8\textendash S1.12).\cite{Ringsby2021, Fong2020, Hansen1990, Peiris2023, Fang2023, Self2019, Han2017, Cresce2011, Xu2012, Xu2007, Seo2015, Ong2015, Borodin2016, Logan2018, Gering2006, Yao2023, Yang2010, Berhaut2019, Fong2019, Solano2013, LiyanaArachchi2018, Forsyth2016, Orbakh1999, Nagl2023, Kar2014, Lee2020, Gering2017, Ding2001, Bjerrum1926, Marcus2006}

\subsection*{Solvation Shell Radius and Electrolyte Viscosity}
\begin{figure}[b!]
\begin{center}
\includegraphics[width=\linewidth]{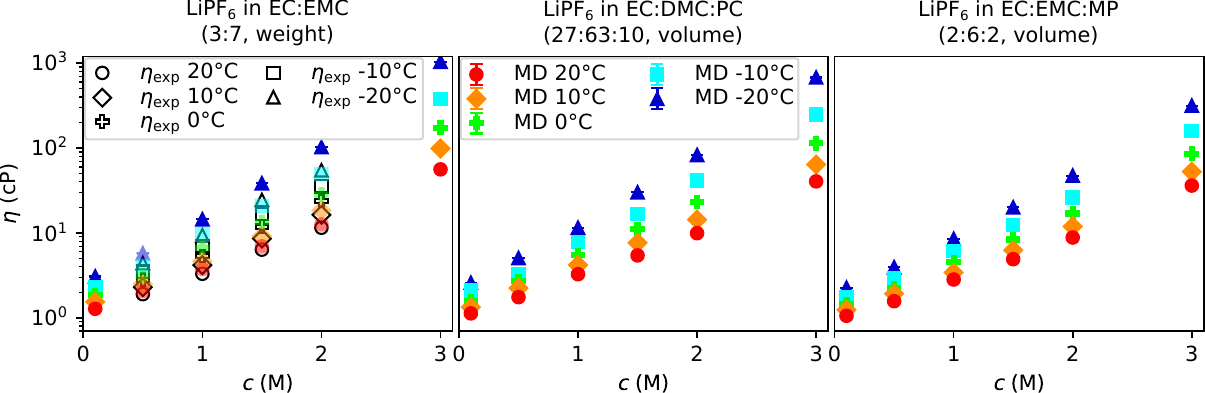}
\caption{Calculated viscosity $\eta$ using MD simulations. The experimental data points $\eta_\mathrm{exp}$ correspond to the Arrhenius fits of the experimental values from Logan et al.\cite{Logan2018} (see SI Figure S13a).}
\label{pic_viscosity}
\end{center}
\end{figure}

The radius of the ion solvation shell $r_\mathrm{s}$ and the electrolyte viscosity $\eta$ directly impact the ionic transport. Assuming Stoke's Law as an approximation for our concentrated electrolytes,\cite{Stokes1851} the dragging force on a solvated ion is directly proportional to both, $r_\mathrm{s}$ and $\eta$. According to this equation, large solvation shell radii and more viscous electrolytes lead to slower movement and lower mobilities of the ions (see SI Section S1.12).\\
While in our MD simulations the composition of the ion solvation shells changes, the corresponding size is independent from temperature and concentration and is similar for our three electrolytes (see SI Figure S9). Consequently, we assume that the solvation shell radius does not contribute to any differences in the transport parameters of our three electrolytes.\\
The viscosity $\eta$ can be approximated using the Stokes-Einstein relation, which connects the diffusion of spherical particles through a liquid to the liquid's viscosity
\begin{equation}
\label{equ_viscosity}
\eta = \frac{k_\mathrm{B}T}{6\pi D^\mathrm{COM}_\mathrm{self}r(c)},
\end{equation}
where $D^\mathrm{COM}_\mathrm{self}$ denotes the self-diffusion coefficient in the COM frame and $r(c)$ the radius of the particles. To apply the Stokes-Einstein relation in our work, we approximate the three component mixture of our electrolytes as a single component liquid, consisting of identical, spherical particles. The corresponding effective self-diffusion coefficient $D^\mathrm{COM}_\mathrm{self}$ of the spherical particles is calculated by composition-weighted averaging the self-diffusion coefficients $D^{i,\mathrm{COM}}_\mathrm{self}$ of the individual electrolyte species $i$ 
\begin{equation}
D^\mathrm{COM}_\mathrm{self}=\sum_{i}^n\omega_i D^{i,\mathrm{COM}}_\mathrm{self},
\end{equation}
where $\omega_i$ denotes the species mass ratio (see SI Eqs. S19\textendash S21).
Additionally, we compare our calculated viscosity values to experimental data to estimate the effective particle radius $r(c)$ (see SI Figure S13). In contrast to the solvation shell radius $r_\mathrm{s}$ of the ions, $r(c)$ represents the average size of all particles involved, including the solvent molecules. Note that Ringsby et al. consider only the self-diffusion coefficients of the solvent species to separate the influence of solvent viscosity and ion association on the transport parameters.\cite{Ringsby2021} For our electrolytes, however, the electrolyte viscosity and the solvent viscosity are almost identical (see SI Figure S15). Therefore, and for the sake of simplicity, we incorporate the self-diffusion coefficients of all species in our calculations. Alternatively, although not done here, the viscosity can also be determined directly from MD simulations using the Green-Kubo relation or inferred from finite size effects.\cite{Jamali2018,Schoen1985,Hess2002}\\
Figure \ref{pic_viscosity} shows the approximated viscosity $\eta$ for all three electrolytes. Our calculated data match the Arrhenius fit of the experimental data $\eta_\mathrm{exp}$ for LiPF$_6$ in EC:EMC (see SI Eq. S22 and SI Table S8) and exhibit for all three electrolytes similar trends: $\eta$ increases with concentration and decreases with temperature. Therefore, the particles experience less drag and enhanced mobility at low concentrations and high temperatures. Overall, LiPF$_6$ in EC:EMC reveals the highest and LiPF$_6$ in EC:EMC:MP the lowest viscosity, suggesting an increasing particle mobility in the same order.
\begin{figure}[tb!]
\begin{center}
\includegraphics[width=\linewidth]{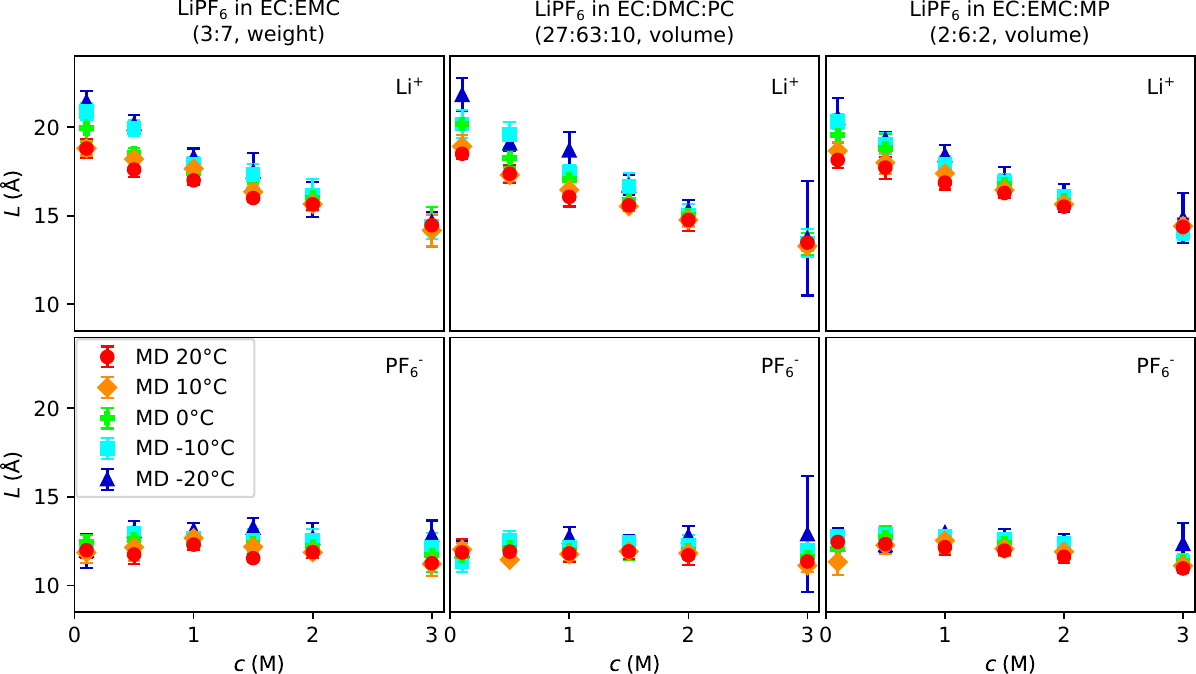}
\caption{Composition-weighted diffusion length $L$ of the solvent molecules, surrounding the Li$^+$ and the PF$_6^-$ ions.}
\label{pic_comp_weighted_diff_length}
\end{center}
\end{figure}

\subsection*{Diffusion Mechanism}
\begin{figure}[tb]
\begin{center}
\includegraphics[width=\linewidth]{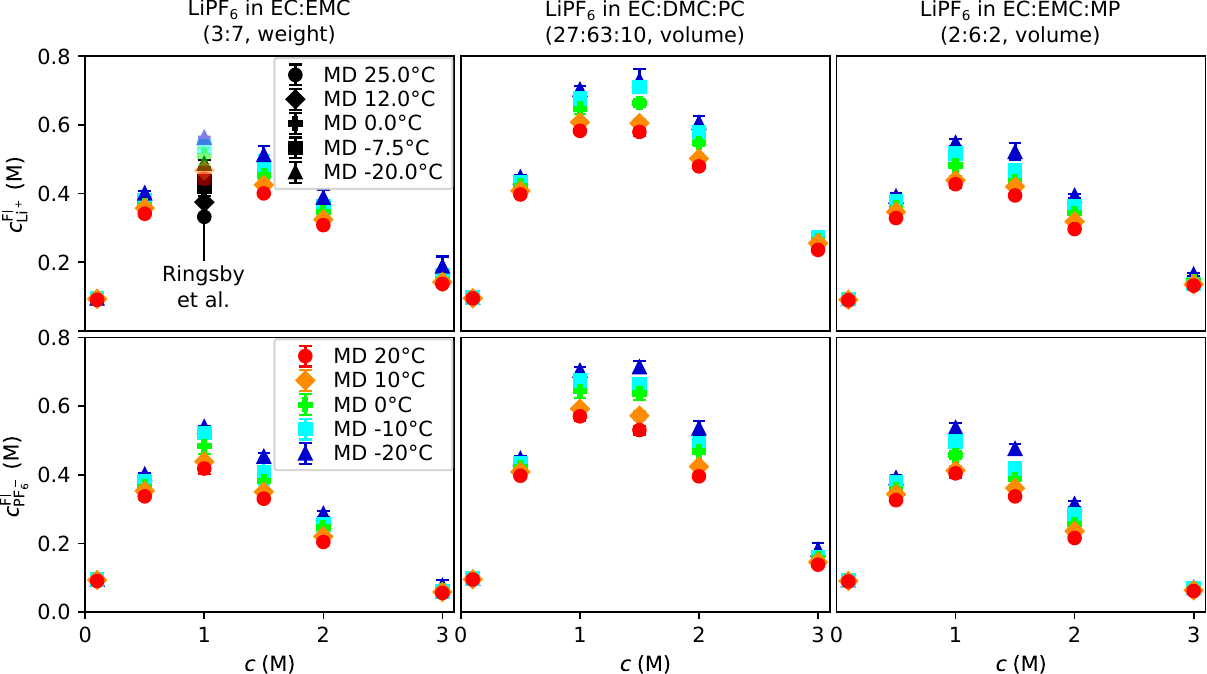}
\caption{Concentrations of free Li$^+$ and PF$_6^-$ ions calculated by MD simulations. For 1\,M LiPF$_6$ in EC:EMC, our FI concentrations match the data from Ringsby et al.,\cite{Ringsby2021} using a scaling factor of $\zeta=0.8$ in their MD simulations.}
\label{pic_free_ions}
\end{center}
\end{figure}
The interaction between the solvent molecules and ions dictates the diffusion mechanism of the ions. Vehicular type diffusion describes the motion of ions surrounded by a stable solvation shell. The individual solvent molecules of species $i$ within the solvation shell accompany the ions over long diffusion lengths $L_i$ before they are substituted by other molecules or ions. In contrast, structural type diffusion reveals short $L_i$, resulting in an ion hopping process with a rapid exchange of solvation shell molecules.\cite{Ringsby2021,Self2019,Fong2019} This type of diffusion relates to faster movement speeds of the ions.\cite{Ringsby2021,LiyanaArachchi2018,Forsyth2016,Gering2017}\\
In order to estimate which diffusion mechanism is prevalent in our three electrolytes, we calculate the corresponding composition-weighted diffusion lengths $L=\sum_{i}^n\omega_i L_i$ (for $L_i$, see SI Figures S19, S20) of the solvent molecules surrounding the Li$^+$ and the PF$_6^-$ ions. $L$ shows for all electrolytes similar results (see Figure \ref{pic_comp_weighted_diff_length}). While the solvent molecules of the Li$^+$ ion solvation shell exhibit decreasing diffusion lengths with concentration and temperature, the molecules surrounding the PF$_6^-$ ion reveal fairly constant and overall lower values for $L$. This indicates, that the solvent molecules are more strongly bound to the Li$^+$ ion,\cite{Ong2015} even at the highest temperatures and concentrations. Consequently, the Li$^+$ ions exhibit a more vehicular type of transport than the PF$_6^-$ ions.\\
Note that the diffusion mechanism can also be analyzed using different methods, as for instance shown in Refs. \citenum{Mistry2023} and \citenum{Zhang2021}.

\subsection*{Ion Association}
Ion associations are defined by the particles occupying the solvation shell. An ion association is formed, if at least one additional ion populates the same solvation shell. This reduces the amount of free ions (FI) in the electrolyte, influencing the transport properties of the electrolyte. Additionally,  the formation of charged associations like negative or positive triple ions (NTI, PTI) can cause the involved Li$^+$ and PF$_6^-$ ions to migrate contrary to the expected direction. This was experimentally demonstrated at very high ion concentrations, namely for Li$^+$ ions in an ionic-liquid-based electrolyte.\cite{Gouverneur2018, Brinkkoetter2018}\\
In order to calculate the amount of free Li$^+$ and PF$_6^-$ ions, we analyze the population of the corresponding solvation shells (see SI Figure S21). Figure \ref{pic_free_ions} shows that both ion species exhibit for all three electrolytes similar behavior: the fraction of FI decreases with temperature and concentration. This leads to a bell shaped behavior of the total FI concentrations, where the highest values are observed at low temperatures and intermediate salt concentrations. At elevated salt concentrations, the amount of FI is slightly lower for the PF$_6^-$ ions compared to the Li$^+$ ions. Our obtained FI concentrations for 1\,M LiPF$_6$ in EC:EMC match the data from Ringsby et al.,\cite{Ringsby2021} using a scaling factor of $\zeta=0.8$ in their MD simulations. Further analysis of the data can be found in the supplementary information (see SI Section S1.11).\\
Overall, LiPF$_6$ in EC:DMC:PC exhibits the highest FI concentrations. This could be due to its elevated cyclic carbonate concentration (see SI Table S1), as cyclic carbonates improve the solubility of the ions.\cite{Borodin2016} LiPF$_6$ in EC:EMC and LiPF$_6$ in EC:EMC:MP show lower, almost identical amounts of FIs. The amounts of NTI and PTI are for all electrolytes negligibly small (see SI Figure S21). 
 
\subsection*{Impact on the Transport Parameters}
Having identified the viscosity, the diffusion mechanism and the formation of ion associations for our three electrolytes, we now estimate their impact on the conductivity, the transference number and the diffusion coefficient.\\ 
The conductivity of an electrolyte depends on the concentration, association and mobility of the ions. Electrolytes containing highly mobile and charged ions reveal increased conductivity values. Here, we choose the free ions (FI, see Figure \ref{pic_free_ions}) as descriptor to analyze the impact of ion concentration and ion association on the ion conductivity. Additionally, we calculate the electrophoretic mobilities (see SI Figure S22), depending on the prevalent diffusion mechanism of the ions and especially on the viscosity $\eta$ (see SI Section S1.12).\\
The combination of the monotonic electrophoretic mobilities and the FI concentrations leads to the conductivities shown in Figure \ref{pic_cond_VOL}. At low salt concentrations, high mobilities are countered by low FI concentrations, resulting in reduced conductivities. At intermediate concentrations, both quantities are sufficiently large to allow for maximum conductivities before they simultaneously decrease at elevated salt concentrations. The temperature dependencies of the mobilities and the FI concentrations are opposite: While the mobilities increase with temperature, the FI concentrations decrease. However, the impact of the mobilities on the conductivity seems to dominate the impact of the FI concentration for our electrolytes, as we observe increasing conductivity trends with temperature.\\
LiPF$_6$ in EC:DMC:PC exhibits the highest electrophoretic mobilities and FI concentrations, and therefore it takes the overall highest conductivity values. On the other hand, LiPF$_6$ in EC:EMC:MP shows lower mobilities and FI concentrations that leads to slightly decreased conductivities. Nevertheless, due to the less prominent decrease of the mobilities with decreasing temperature, LiPF$_6$ in EC:EMC:MP surpasses the conductivity in EC:DMC:PC at $-20$\,\textdegree C. Finally, since LiPF$_6$ in EC:EMC contains similar FI concentrations as in EC:EMC:MP but simultaneously exhibits the lowest electrophoretic mobilities, it becomes the least conductive of this study.\\
The transference number $t^\mathrm{COM}_+$ depends on differences in the formation of ion associations and the corresponding electrophoretic mobilities of anions and cations. Faster electrophoretic Li$^+$ ion mobilities and elevated free Li$^+$ concentrations lead to increased values of $t^\mathrm{COM}_+$. As evident in Figure \ref{pic_comp_weighted_diff_length}, the Li$^+$ ions show decreasing diffusion lengths $L$ with increasing salt concentration for all three electrolytes, indicating a transition from vehicular to a more structural type of diffusion mechanism. Additionally, using again the FI concentration to estimate the impact of ion association on $t^\mathrm{COM}_+$ reveals elevated FI concentrations of lithium cations compared to the PF$_6^-$ ions (see Figure \ref{pic_free_ions}). This suggests that more anions participate in larger ion associations. Combining both results leads to a slight increase of $t^\mathrm{COM}_+$ with increasing concentration and even to $t^\mathrm{COM}_+ \approx t^\mathrm{COM}_-$ at $c=3.0\,$M (see SI Figure S3). The PF$_6^-$ ions exhibit a diffusion process that is overall more structural, which explains the low transference numbers for $c<3.0\,$M. \\
The diffusion coefficient $D^\mathrm{COM}_\pm$ follows similar dependencies as the conductivity. According to the Einstein-Smoluchowski relation, the diffusion coefficient for dilute electrolytes (electrolytes without ion-ion correlations) is directly proportional to the electrophoretic mobility. Therefore, elevated mobilities result in faster diffusion processes. As mentioned above, the mobilities depend on the prevalent diffusion mechanism and more prominently on the viscosity. Since the viscosity increases with concentration and decreases with temperature, $D_\pm^\mathrm{COM}$ is expected to exhibit contrary trends. However, while the diffusion coefficient in fact increases with temperature, it still shows elevated values at high concentrations. This is due to the fact that $\mathit{TDF}^\mathrm{COM}$ increases with concentration and therefore, it compensates the loss in mobility (see SI Figure S3). Interestingly, $\mathit{TDF}^\mathrm{COM}$ has the highest value for the electrolyte with the lowest viscosity and vice versa. Therefore, the diffusion coefficients are similar for all three systems. 

\section*{Conclusion}
In this work, we combined MD simulations with measurements on conductivity  and concentration cells in order to determine the electrolyte parameters of LiPF$_6$ in solutions of EC:EMC (3:7, weight), EC:DMC:PC (27:63:10, volume), and EC:EMC:MP (2:6:2, volume) at $-20$\,\textdegree C$\leq T\leq$20\,\textdegree C and 0.1\,M$\leq c\leq$3.0\,M. In contrast to commonly used polarization experiments, our method avoids the error-prone usage of Li metal electrodes with carbonate-based electrolytes.\\
For the MD simulations, we employed the non-polarizable OPLS force fields with a constant scaling factor $\zeta$, reducing the effective charges of the involved Li$^+$ and PF$_6^-$ ions. To calibrate $\zeta$ we compared the simulated conductivities to experimental data. This allows for a more precise determination of the transference number $t_+$. The combination of the simulated transference number with experimental data from concentration cells enabled the deconvolution of the thermodynamic factor $\mathit{TDF}$ and the calculation of the diffusion coefficient $D_\pm$. For the purpose of validating our MD simulations, we employed eNMR measurements and determined $t_+$ for LiPF$_6$ in EC:EMC and in EC:EMC:MP. For both electrolytes, the experimental data show strong agreement with the calculations. However, comparing our data to the electrolyte parameters determined by polarization experiments using Li metal electrodes yielded significant deviations, especially for the salt diffusion coefficient $D_\pm$. These discrepancies can be attributed to detrimental interphasial effects between the Li metal electrodes and the carbonate-based electrolytes, which make the evaluation of the polarization experiments unfeasible.\cite{Bergstrom2021, Lehnert2024,Talian2019}\\
Additionally, we evaluated the electrolyte viscosity, the diffusion mechanisms, and the formation of ion associations and qualitatively estimated their impact on the individual transport parameters.

\section*{Author Contributions}
Lukas Lehnert contributed to the conceptualization of the paper, wrote the text and created the figures of the manuscript and the supplementary information. He mixed the examined electrolytes and performed the concentration cell measurements, implemented and parametrized the MD simulations and analyzed and interpreted the obtained data. Martin Lorenz and Prof. Dr. Monika Schönhoff performed and analyzed the eNMR data of the electrolytes and wrote the corresponding theory and experimental sections. Dr. Maria Fernanda Juarez supervised the MD simulations and reviewed the whole article. Dr. Max Schammer and Dr. Maryam Nojabaee supervised and reviewed the employed continuum electrolyte theory and the concentration cell measurements, respectively. Prof. Dr. Birger Horstmann supervised the whole work, contributed to the conceptualization of the paper, reviewed the article, and acquired the funding.

\section*{Acknowledgements}
We gratefully acknowledge the support by the project DeepBat, a collaboration between the National Aeronautics and Space Administration – Jet Propulsion Laboratory (NASA – JPL) and the German Aerospace Center (DLR), and appreciate the support by the state of Baden-Württemberg through bwHPC and the German Research Foundation (DFG) through grant no INST 40/575-1 FUGG (JUSTUS 2 cluster). We acknowledge funding of the NMR spectrometer by the DFG (German Science Foundation) via project ID 452849940. We express our gratitude to the Center for Electrochemical Energy Storage Ulm \& Karlsruhe (CELEST) and gratefully acknowledge the glass-blowing of the University of Stuttgart for manufacturing the concentration cells. We thank Prof. Dr. Kara Fong and Alexandra Ringsby for their helpful and inspiring correspondence and are grateful to Prof. Dr. Agilio Padua for providing the fftool on Github, which facilitated the preparation of the input files for the MD simulations.

\section*{Conflict of Interest}

The authors declare no conflict of interest.




\setlength{\bibsep}{0.0cm}
\bibliographystyle{bib_style}
\bibliography{example_refs}

\end{document}